\documentclass[11pt]{article}
\usepackage[preprint]{acl}
\usepackage{times}
\usepackage[T1]{fontenc}
\usepackage[utf8]{inputenc}
\usepackage{microtype}
\usepackage{amsmath,amssymb,booktabs,graphicx,array,tabularx,placeins}
\title{Verification Allocation in Stylized LLM Agent Pipelines under Hidden Tool Degradation: A Discrete-Time Mean-Field Analysis}
\author{Shereen Anand\\PES University\\\texttt{pes1202201328@pesu.pes.edu}\And Chandramani Singh
\\Indian Institute of Science\\\texttt{chandra@iisc.ac.in}}
\date{}
\begin{document}
\maketitle
\begin{abstract}
We study a stylized verification-allocation model for role-structured pipelines with binary inherited corruption and hidden tool degradation. Standard contraction arguments give serial and multi-role fixed points and a finite-population deviation bound under independence. When verification actions affect neither future errors nor observations, the belief-state choice is myopic. We characterize a two-action threshold under an explicit monotonicity condition and a restricted four-action interval/tail policy when generation rates are equal and only one check degrades. Synthetic experiments check these model-specific results and find counterexamples outside the restriction. This is conditional model analysis, not a general mean-field-game result or empirical evidence about deployed LLM systems.
\end{abstract}
\section{Introduction}\footnote{\href{https://github.com/Shereen20112003/verification-allocation-hidden-tool-degradation}{GitHub}.}
Multi-agent language-model systems decompose a task into intermediate artifacts passed between specialized roles. Errors can propagate, and a verifier can share the tool degradation it is meant to detect. We study a stylized model with binary inherited corruption, two hidden tool regimes, and priced verification choices.

The question is how to allocate differently priced checks when one check's reliability is hidden and changes over time. A runner-based test may be cheap when healthy but unreliable under shared-tool degradation; an independent reviewer may be robust but expensive. A noisy mismatch signal provides indirect evidence of tool health.

Prior work documents failure and attribution in LLM-agent teams \citep{cemri-etal-2025,zhang-etal-2025} and role-structured frameworks \citep{wu-etal-2023,hong-etal-2023,qian-etal-2023}. Adaptive verification has been studied for reasoning and factual claims \citep{wan-etal-2024,zhong-etal-2025,hassan-etal-2026}; budgeted verification and corrector placement are closer allocation problems \citep{xue-etal-2026,itkin-2026}, while tool-retrieval work adapts to execution reliability \citep{chen-liao-2026}. Our decision object is different: checks for inherited binary errors under a hidden shared regime. We assume actions affect neither future errors nor signals, making the policy myopic. The analysis uses standard population-process and POMDP ideas \citep{lovejoy-1987,watkins-etal-2017,sridhar-kar-2023}. Related mean-field-game work studies strategic equilibria \citep{wang-chakraborty-2025,lasry-lions-2007,carmona-delarue-2018}; here agents do not optimize strategically and actions do not alter future states. These are model-specific results, not new general mean-field-game or closed-loop control theory. A targeted search found no exact match, but does not establish priority.

\paragraph{Contributions.}
\textbf{Propagation.} For this binary-error model, we analyze serial and role-structured recursions, their fixed points, and a finite-population deviation bound under explicit independence assumptions.

\textbf{Allocation.} Under action-independent transitions and signals, we characterize pointwise optimal choices, a two-action threshold under a stated derivative condition, and a four-level interval/tail structure when generation rates are equal and only the runner degrades.

\textbf{Experiments.} We implement and run a deterministic-seed CPU suite (E1-E6) with Monte Carlo replications, topology and finite-size ablations, filter stress tests, equal-budget comparisons, and randomized structural checks. The outputs are synthetic model checks; there are no calls to LLMs and no real-framework data.

The experiments are internal consistency checks, not evaluations of deployed systems.

\section{Model and assumptions}
Time is discrete. A unit (e.g., a method, section, or subtask) has state x\_t in \{0,1\}, where 1 denotes corrupted. The environment R\_t in \{N,D\} is a two-state Markov chain with transition probabilities P(D|N)=alpha and P(N|D)=beta. The regime is not observed directly.

At each step, the controller chooses a verification level v. A unit fails generation with probability g\_R. If its input is corrupted, verification catches it with probability rho\_v(R); otherwise the corruption is inherited. Conditional on the regime, input state, and level, generation failure and a missed catch are independent. For a single serial chain with input corruption probability m\_t:

\begin{equation}
m_{t+1}=g+(1-g)(1-\rho)m_t
\end{equation}
A role-structured system has K roles. The input role for a role-k unit is sampled according to row-stochastic matrix P. With role failure vector g and role catch rates rho-bar, the mean-field update is

\begin{equation}
\begin{aligned}m(t+1)&=g+A m(t),\\A&=\operatorname{diag}((1-g_k)(1-\bar{\rho}_k))P\end{aligned}
\end{equation}
For hidden-regime control, the controller observes a binary signal s\_t with P(s=1|D)=p and P(s=1|N)=q, p>q. Let b\_t=P(R\_t=D|s\_1:t). A level costs kappa\_v and incurs penalty lambda for corruption. With input corruption probability mu, the one-step expected cost is

\begin{equation}
c(b,v)=\kappa_v+\lambda[g(b)+(1-g(b))(1-\rho_v(b))\mu]
\end{equation}
\subsection{Assumptions and scope}
The four levels are: no check (v=0), a regime-independent reviewer (v=1), a runner-based check (v=2), and a robust re-sampling or self-consistency check (v=3). Their costs satisfy 0=kappa\_0 < kappa\_2 < kappa\_1 < kappa\_3; nominal catches satisfy 0=rho\_0 < rho\_1 <= rho\_2 < rho\_3 <= 1. Only level 2 degrades: rho\_2(D)=delta rho\_2(N), with 0<=delta<1. The input corruption probability mu is treated as a parameter in the allocation analysis. The robust re-sampling interpretation is related to self-consistency methods for language-model reasoning \citep{wang-etal-2022}.

These assumptions intentionally exclude several realities: partially wrong artifacts, dependence between agents sharing a model, endogenous or adversarial signals, and verification actions that alter the future state. Results that rely on conditional independence do not extend automatically to correlated failures.

A notable modeling distinction is between inherited corruption and generation failure. Verification only intercepts the former; it cannot repair a newly generated error. Hence perfect verification does not drive the corruption rate to zero when g>0.

\section{Serial propagation}
The scalar recursion is affine. Its slope is a=(1-g)(1-rho), which is strictly below one whenever g>0 or rho>0. Solving m*=g+a m* gives the following exact result.

\begin{equation}
m^*=\frac{g}{g+\rho-g\rho}
\end{equation}
\textbf{Proposition 1 (serial fixed point).} For positive g+rho-g rho, the recursion has a unique fixed point m*. Moreover |m\_t-m*| is at most a\textasciicircum{}t |m\_0-m*|. With rho=0 and positive g, m*=1; with rho=1, m*=g.

The fixed point follows by rearranging m*=g+(1-g)(1-rho)m*. The denominator g+rho-g rho is positive under the stated condition. The contraction factor is nonnegative and below one, giving geometric convergence. Differentiating the fixed point shows it increases with g and decreases with rho for interior parameter values.

The no-verification endpoint is a qualitative warning: in an infinite serial chain, any positive independent generation-failure rate eventually drives the stationary corruption fraction to one. A nonzero catch rate prevents total collapse in this stylized model, but the required catch rate does not depend on the chain length. This is a steady-state statement, not a claim about finite-horizon task accuracy.

\subsection{Static verification choice}
For a fixed regime and a given input corruption probability mu, the per-unit expected cost of level v is kappa\_v + lambda[g+(1-g)(1-rho\_v)mu]. Comparing adjacent levels gives the exact incremental rule:

\begin{equation}
\kappa_v-\kappa_{v-1}\leq\lambda(1-g)\mu(\rho_v-\rho_{v-1})
\end{equation}
In the self-dependent serial chain, the input corruption probability is itself the stationary value under the chosen level. The relevant comparison is then between total costs evaluated at the two fixed points, m*\_v=g/(g+rho\_v-g rho\_v). One cannot substitute the one-step incremental formula without holding mu fixed.

\subsection{Interpretation}
The model separates two economic effects: a check costs tokens regardless of whether it is needed, while it reduces only the inherited-error component. When generation failure dominates, even a perfect check has limited value. When inherited corruption is large, a cheap, reliable check can be worth more than a robust but expensive alternative.

\subsection{Numerical check: E1}
E1 evaluates the stationary formula on 72 combinations: eight generation rates from 0.02 to 0.30, nine catch rates from 0 to 0.95, and 200,000 Monte Carlo units per combination. To avoid a slow burn-in at the no-check edge, each case starts from the analytic stationary marginal and applies one model transition. This tests invariance of the predicted marginal; it does not estimate convergence speed.

The largest absolute empirical discrepancy is 0.0029. All 64 positive-catch cells fall within three nominal Monte Carlo standard errors. At rho=0, every simulated chain is corrupted after the transition and the empirical fraction is exactly one.

This design is deliberately transparent: the experimental process is the same process used in the theorem. Agreement can detect implementation or algebra errors, but it does not validate the binary-state model against empirical LLM behavior.

\begin{figure}[t]
\centering
\includegraphics[width=\columnwidth]{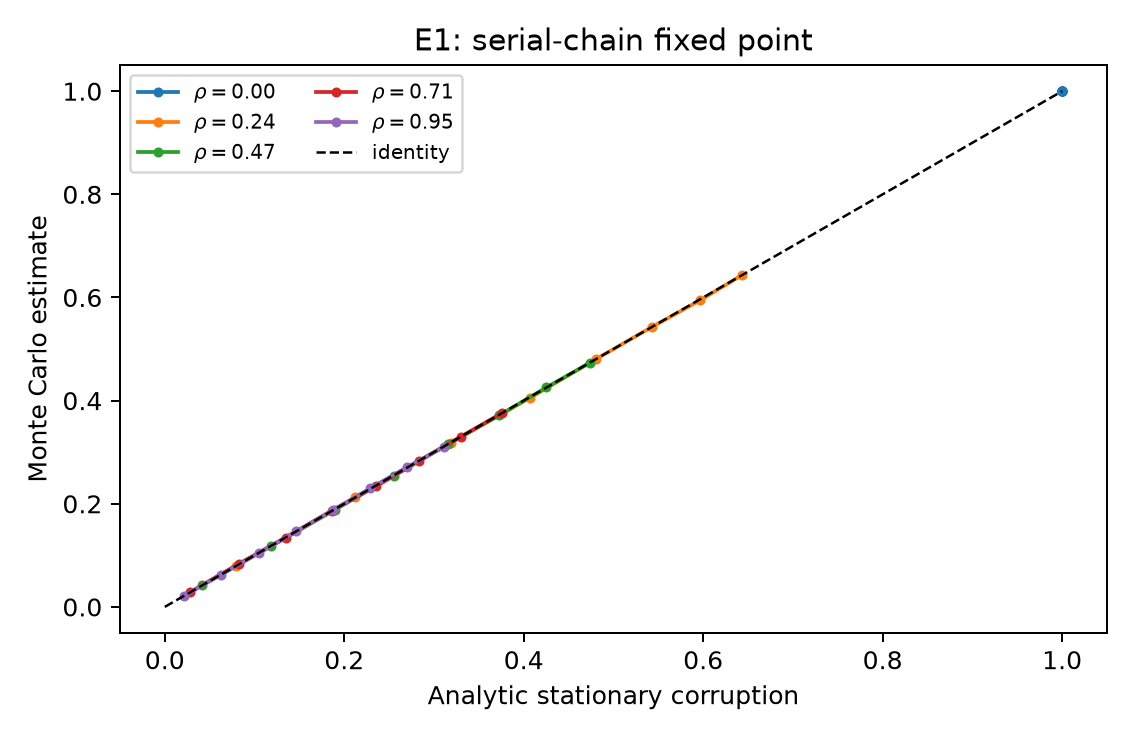}
\caption{E1 stationary marginal estimates versus the analytic fixed point.}
\label{fig:paper-1}
\end{figure}
\section{Role-structured teams}
Let g\_k be the generation-failure rate of role k and let rho-bar\_k be its effective catch rate. Role k inherits a corrupted input with probability (P m)\_k. The mean-field update is affine with A=diag((1-g\_k)(1-rho-bar\_k))P.

\begin{equation}
m(t+1)=g+A m(t)
\end{equation}
\textbf{Theorem 2 (fixed point and contraction).} If u=max\_k (1-g\_k)(1-rho-bar\_k)<1, then the spectral radius of A is at most u, the unique fixed point is m*=(I-A)\textasciicircum{}(-1)g, and ||m(t)-m*||\_infinity <= u\textasciicircum{}t ||m(0)-m*||\_infinity.

Each row of A is nonnegative and has sum at most u, so its infinity norm is at most u. The Neumann series gives (I-A)\textasciicircum{}(-1)=sum\_(j>=0) A\textasciicircum{}j. The contraction bound follows by subtracting the fixed-point equation from the recursion. The condition is sufficient, not necessary; a zero-failure, zero-verification role can require a separate invariant-subspace analysis.

\subsection{Heterogeneity sensitivity}
For two systems with matrices A and A' and failure vectors g and g', assume both infinity norms are at most u<1. The resolvent identity yields

\begin{equation}
\begin{aligned}\|m^*-m^{*\prime}\|_\infty&\leq\frac{1}{1-u}\bigl(\|g-g'\|_\infty\\&\qquad+\|A-A'\|_\infty\|m^{*\prime}\|_\infty\bigr)\end{aligned}
\end{equation}
The bound separates parameter mismatch in the local failure rates from mismatch in the interaction structure. The factor 1/(1-u) makes sensitivity large near weak contraction. It is an upper bound and can be loose; it does not assert that every role is equally sensitive.

\subsection{E2: topology experiment}
E2 evaluates K in \{2,3,5\} under a serial predecessor chain, a random row-stochastic graph, and a hub-and-spoke graph. Role failures and catches vary by role. For each topology, the script solves (I-A)m*=g and simulates synchronous populations with independent binomial role updates. It compares 4,000 agents per role with an imbalanced allocation of 1,000, 2,000, ... agents.

Across 60 role/topology/population rows, mean absolute error is 0.00154 and the largest discrepancy is 0.01038. The time series uses a finite burn-in and sampling window, so these are Monte Carlo approximations rather than exact equilibrium samples.

\section{Finite teams and concentration}
Consider N\_k agents in role k. Each agent independently samples one input from the prescribed source role and independently generates or catches errors conditional on that input. Let m-hat\_k\textasciicircum{}N(t) be the empirical corrupted fraction and m\_k(t) the mean-field trajectory initialized at the same population mean.

\textbf{Theorem 3 (finite-team deviation).} With u=max\_k (1-g\_k)(1-rho-bar\_k)<1 and N\_min=min\_k N\_k,

\begin{equation}
\begin{aligned}\mathbb{E}|\hat m_k^N(t)-m_k(t)|&\leq\frac{1-u^t}{2(1-u)\sqrt{N_{\min}}}\\&\leq\frac{1}{2(1-u)\sqrt{N_{\min}}}\end{aligned}
\end{equation}
Conditional on the current populations, the next role-k count is binomial with mean N\_k nu\_k(t). Its expected absolute deviation from its conditional mean is at most 1/(2 sqrt(N\_k)). The conditional mean differs from the deterministic update by at most u times the previous maximum expected deviation. Thus e\_t <= u e\_(t-1)+1/(2 sqrt(N\_min)); summing the geometric series gives the bound.

\subsection{Limitations of the bound}
The theorem requires conditionally independent agents and independent source sampling with replacement. Shared model errors can produce cross-agent dependence and invalidate the binomial step. The constant also deteriorates as u approaches one. We therefore report both empirical finite-size scaling and the stated bound, rather than describing the bound as a tight finite-sample guarantee.

\subsection{E3: finite-size scaling}
E3 uses a three-role cyclic topology, 100 independent population replicas at each of nine population sizes from 250 to 64,000 per role, and 35 synchronous updates. For each role, we average absolute error across replicas and report the maximum of those role-wise means. This statistic is bounded by the theorem's per-role expectation bound. A least-squares line is fitted in log error versus log N.

For fast mixing (u=0.368), the fitted slope is -0.506; for slow mixing (u=0.960), it is -0.495. No mean-over-replica row exceeds the theorem's conservative upper bound. These slopes are single-seed simulation summaries, not uncertainty-adjusted estimates of a universal exponent.

\section{Hidden regime and belief-state control}
Let b=P(R=D) be the filtered probability of the degraded regime. The prediction step is b-bar=b(1-beta)+(1-b)alpha. Given signal s, Bayes' rule gives b'=ell\_D(s)b-bar/[ell\_D(s)b-bar+ell\_N(s)(1-b-bar)]. Here ell\_D(1)=p, ell\_N(1)=q, with complementary likelihoods for signal zero.

\textbf{Proposition 4 (exact myopia).} If the chosen level affects neither the regime transition nor the signal law and affects only the current stage cost, every optimal stationary action minimizes c(b,v) pointwise. Discounting does not change the minimizing action.

In the Bellman equation, the expected continuation value depends on b and the next signal but not on v. It is therefore an action-independent additive term. The result is exact, but narrow: once verification changes future corruption, the continuation value depends on the action and myopia no longer follows.

\subsection{Two-level threshold}
Let fragile level F have catch rho\_F in N and delta rho\_F in D; let robust level R have constant catch rho\_R>rho\_F and a larger cost. Define h(b)=(1-g(b))(rho\_R-rho\_F(1-(1-delta)b)). The incremental robust-minus-fragile cost is (kappa\_R-kappa\_F)-lambda mu h(b).

\textbf{Theorem 5 (threshold condition).} Assume g\_D>=g\_N. Then h is nondecreasing exactly when 

\begin{equation}
(1-\delta)\rho_F(1-g_D)\geq(g_D-g_N)(\rho_R-\delta\rho_F)
\end{equation}
Under this condition, the optimal action switches at most once from F to R as b increases. It is an interior threshold when the incremental robust cost lies strictly between lambda mu h(0) and lambda mu h(1). If g\_D=g\_N, the condition holds automatically. If g\_D is below g\_N, h is also increasing for the parameterization above, but this case is not needed for the degradation interpretation.

The restriction g\_D>=g\_N is stated explicitly here to align the endpoint derivative argument with the intended direction of degradation. Randomized checks in E6 cover both signs of g\_D-g\_N and compare the derivative condition directly with grid monotonicity.

\subsection{Filtering experiment}
E4 computes the 4-level myopic policy over 1,001 belief values and evaluates post-change detection delay. The regime is assumed to have switched to D at time zero for the delay calculation; each trial then draws D-regime signals until the posterior crosses 0.9. This is a change-detection stress test, not a full closed-loop Markov simulation.

\section{Four-level structure and observability}
Suppose g\_N=g\_D=g, so degradation changes tool quality but not the generation failure rate. Levels 0, 1, and 3 then have belief-independent costs. The runner's level-2 catch rate rho\_2(b)=rho\_2[1-(1-delta)b] decreases with b, so its cost is nondecreasing. Its set of beliefs at which it beats the best belief-independent level is an initial interval, possibly empty.

\textbf{Proposition 6 (four-level initial interval and tail).} Under g\_N=g\_D, the optimal action is level 2 on an initial belief interval [0,tau], if that interval is nonempty, and thereafter is the constant minimizer among \{0,1,3\} at b=1. The tail can be no verification: a robust check is used only if its reduced expected error at full degradation belief compensates for its cost.

\begin{equation}
\kappa_R\leq\lambda\mu(1-g)\rho_R
\end{equation}
The result is a statement about one-stage expected cost under the specified action-independent information process. It is not a generic theorem for multi-level degradation models. When g\_D differs from g\_N, the costs of levels 0, 1, and 3 also vary with belief and multiple policy switches can occur.

\subsection{Value of observing the regime}
Let f\_v(R) be the expected cost of level v in regime R. The one-step value of observing R directly rather than acting on belief b is

\begin{equation}
\begin{aligned}\mathrm{VoO}(b)&=\min_v[bf_v(D)+(1-b)f_v(N)]\\&\quad-[b\min_vf_v(D)+(1-b)\min_vf_v(N)]\end{aligned}
\end{equation}
The value is nonnegative: choosing before observing cannot improve on the regime-informed choice. It is zero at a given belief when a common action is optimal across the regimes with positive belief mass. It is zero for all beliefs exactly when some level is optimal in both regimes.

\subsection{Main empirical findings}
E4's four-level grid check finds the predicted initial-interval/constant-tail form in all five signal settings; the policy itself does not depend on signal quality under exact myopia. Mean post-change delay falls from 5.065 steps at p=0.30 to 1.030 at p=0.99 (q=0.02; 200 trials per setting).

E5 shows the intended conditional pattern: adaptive allocation improves on the best static mixture at matched average spend in the informative, regime-discriminating scenarios; it yields exactly zero measured improvement when p=q or when the same level is optimal in both regimes. These are constructed synthetic scenarios, not evidence of deployment gains.

\section{Main results and limitations}
The primary quantitative checks are summarized in Table 1. E1-E3 compare simulation with explicit recursions; E4 checks belief-policy structure and filter response; E5 compares information-adaptive and static policies under a matched verification-spend budget; E6 stress-tests structural claims over randomized parameter sets.

\begin{table}[t]
\centering
\small
\setlength{\tabcolsep}{2pt}
\begin{tabularx}{\columnwidth}{>{\hsize=0.4800\hsize\raggedright\arraybackslash}X>{\hsize=1.0200\hsize\raggedright\arraybackslash}X>{\hsize=1.5000\hsize\raggedright\arraybackslash}X}
\toprule
\textbf{Exp.} & \textbf{Scope} & \textbf{Measured result} \\
\midrule
E1 & 72 chain settings; 200k units each & max |error| = 0.0029; 64/64 positive-rho cells within 3 MC SE \\
E2 & 60 role rows; 3 topologies & mean |error| = 0.00154; max = 0.01038 \\
E3 & 2 mixing rates; 9 N; 100 reps & log-log slopes -0.506, -0.495; 0 mean-bound violations \\
E4 & 5 signal qualities; 1001 beliefs & 5/5 policy grids have predicted form \\
E5 & 6 synthetic scenarios; 300k units & gain 0 to 0.0452 at matched spend \\
E6 & 400 two-level; 100 + 500 four-level & 400/400 derivative checks agree; 100/100 equal-g form; 9/500 general-g counterexamples \\
E7 & 9 hosted-LLM prompts; 4 paired tasks & 4/4 baseline and self-check; verifier TP=5, TN=3 (8 claims) \\
\bottomrule
\end{tabularx}
\caption{Main measured checks (fixed seed 20261101).}
\label{tab:main}
\end{table}
\subsection{What these results establish}
Under the implemented model, the finite simulations are consistent with the stated stationary and mean-field calculations, show the expected N\textasciicircum{}(-1/2) empirical rate, and reproduce the restricted threshold and four-level structures. The E5 examples demonstrate that informative signals can be useful when the best action differs by regime. The E6 counterexamples reinforce that the equal-generation premise is substantive. In a separate exploratory hosted-LLM pilot, both answer conditions passed 4/4 tasks; the single verifier batch labeled 5/5 true and 3/3 false claims correctly.

\subsection{What these results do not establish}
The synthetic experiments do not estimate deployed error rates or calibrate the hidden-regime model. E7 is only four paired arithmetic, data, and code tasks plus one eight-claim batch; it has no tool-degradation intervention, multi-agent workflow, or statistical power. It cannot establish a self-check benefit or validate the proposed allocation policy. The synthetic budget comparison still uses hand-set costs and catches.

The policy is myopic because actions do not affect future corruption or observations. A policy that controls future team state, learns verifier reliability online, or accounts for delayed consequences requires a different state model and dynamic program. We do not present the continuous-time mean-field game results of Wang and Chakraborty as results of this paper.

\subsection{Reproducibility}
The attached experiments.py writes CSV summaries and figures under results/. The default seed is 20261101. The environment used for this run was Python 3.12.10 and NumPy 2.5.4 on Windows 11; the released run completed in about 68 seconds, including plot creation. The PDF and source identify all randomization and sample counts. Re-running with the same environment and seed should reproduce the tables.

\FloatBarrier

\FloatBarrier
\FloatBarrier
\appendix
\setcounter{table}{0}
\setcounter{figure}{0}
\renewcommand{\thetable}{A\arabic{table}}
\renewcommand{\thefigure}{A\arabic{figure}}
\section{Protocol and execution}
The following pages report the complete experiment protocol and ablations. Experiments E1-E6 are CPU-only synthetic simulations with common seed 20261101, using NumPy's default Generator with PCG64. E7 is a separate, small pilot using hosted LLM inference; it is not a framework or multi-agent evaluation.

Experiments E1-E3 test identities derived from the same stochastic model, so their role is falsification and implementation checking, not external validation. E4 tests the Bayesian filter and structural predictions on a belief grid. E5 compares policy classes within the model. E6 uses randomized parameter draws to check conditions and find counterexamples.

\subsection{Reproduction commands}
\begin{equation}
\texttt{python experiments.py}
\end{equation}
The script writes results/summary.json, twelve CSV files, and three PNG figures. The privacy-reviewed E7 prompts, responses, and per-call usage records are available in the linked repository at results/e7\_pilot\_records.json and in the source archive. Repeating E7 requires access to the hosted model service and additional request authorization. Reproducibility is environment-sensitive at final Monte Carlo decimals; formulas and the seed are recorded in output metadata.

\section{E1: serial-chain grid}
The grid uses g in \{0.02, 0.06, ..., 0.30\} and nine evenly spaced rho values from 0 through 0.95. At each point, N=200,000 independent stationary inputs are sampled from Bernoulli(m*), then one generation/catch transition is applied. The empirical output corruption fraction is compared with m*. This is a stationarity test and avoids choosing an arbitrary burn-in.

\begin{figure}[t]
\centering
\includegraphics[width=\columnwidth]{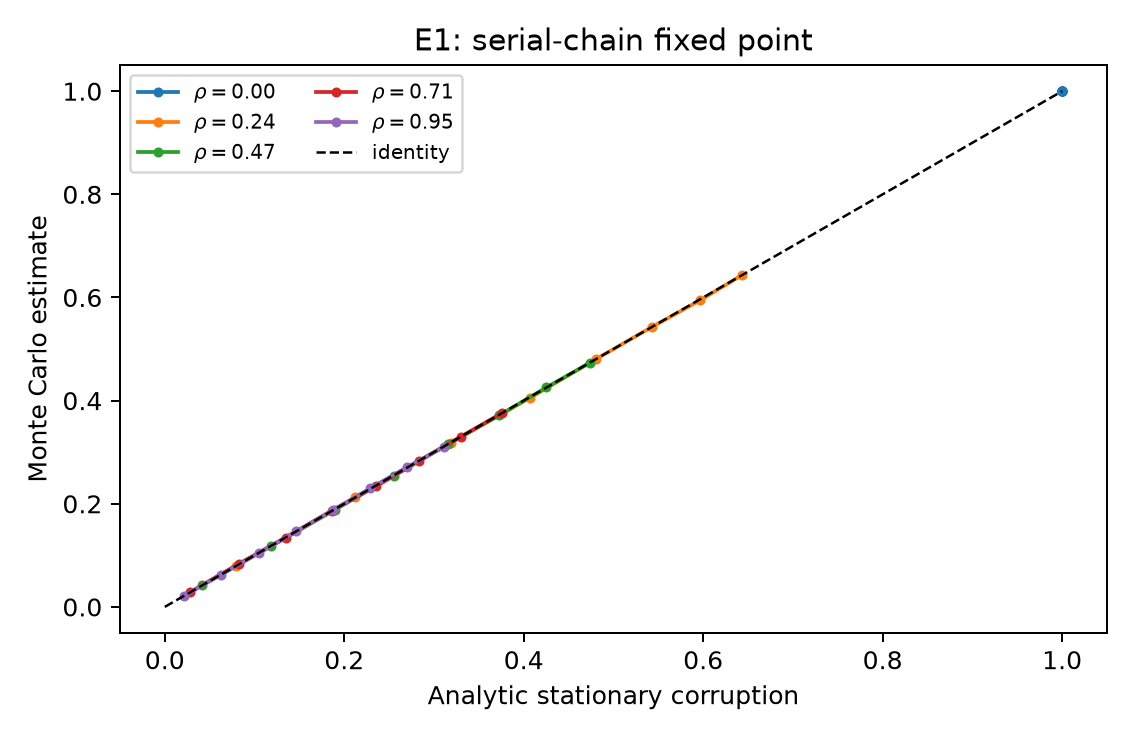}
\caption{Predicted and observed stationary corruption across selected catch rates.}
\label{fig:paper-2}
\end{figure}
Across all 72 grid cells, maximum absolute error is 0.00294. For all 71 cells with positive catch rate, the observed fraction is within three plug-in Monte Carlo standard errors of the prediction. The rho=0 row is degenerate at one after the transition for every g>0. We do not treat these 72 independent parameter cells as 72 independent real-world replications.

\subsection{Chain-length and warm-up ablation}
A separate set starts from an uncorrupted input and simulates 50,000 independent chains to lengths 1, 5, 25, 100, and 500, for g=0.1 and rho in \{0, 0.2, 0.6\}. The measured final fraction is compared with the exact finite-horizon expectation obtained by iterating the scalar recursion. This checks transient behavior rather than only the stationary distribution.

\subsection{Monte Carlo interpretation}
Monte Carlo uncertainty is binomial under this experiment's independence assumptions. The three-standard-error summary is descriptive and does not correct for choosing the grid after observing the results. No parameter tuning was performed to match the observed values.

\section{E1 ablation: chain length}
The chain-length ablation provides a finite-horizon check of the recursion. At rho=0, the state converges toward complete corruption at a rate set by g; positive rho gives a lower fixed point. The gap after t steps contracts by [(1-g)(1-rho)]\textasciicircum{}t, matching the theorem.

\begin{table}[t]
\centering
\small
\setlength{\tabcolsep}{2pt}
\begin{tabularx}{\columnwidth}{>{\hsize=0.8800\hsize\raggedright\arraybackslash}X>{\hsize=1.0400\hsize\raggedright\arraybackslash}X>{\hsize=1.0400\hsize\raggedright\arraybackslash}X>{\hsize=1.0400\hsize\raggedright\arraybackslash}X}
\toprule
\textbf{Length} & \textbf{rho=0} & \textbf{rho=0.2} & \textbf{rho=0.6} \\
\midrule
1 & 0.10034 & 0.09942 & 0.10194 \\
5 & 0.41126 & 0.28808 & 0.15534 \\
25 & 0.92794 & 0.35552 & 0.15506 \\
100 & 0.99998 & 0.3569 & 0.15588 \\
500 & 1.0 & 0.35734 & 0.15812 \\
\bottomrule
\end{tabularx}
\caption{E1 chain-length ablation (g=0.1; Monte Carlo estimates).}
\label{tab:appendix-1}
\end{table}
Each empirical estimate uses 50,000 independent chains. The exact finite-horizon expectation is computed by iterating the recursion from m\_0=0; it is not estimated from the Monte Carlo sample. The two tables separate sampling accuracy from the systematic distance to the stationary value.

\begin{table}[t]
\centering
\small
\setlength{\tabcolsep}{2pt}
\begin{tabularx}{\columnwidth}{>{\hsize=0.8800\hsize\raggedright\arraybackslash}X>{\hsize=1.0400\hsize\raggedright\arraybackslash}X>{\hsize=1.0400\hsize\raggedright\arraybackslash}X>{\hsize=1.0400\hsize\raggedright\arraybackslash}X}
\toprule
\textbf{Length} & \textbf{rho=0} & \textbf{rho=0.2} & \textbf{rho=0.6} \\
\midrule
1 & 0.100 & 0.100 & 0.100 \\
5 & 0.410 & 0.288 & 0.155 \\
25 & 0.928 & 0.357 & 0.156 \\
100 & 1.000 & 0.357 & 0.156 \\
500 & 1.000 & 0.357 & 0.156 \\
\bottomrule
\end{tabularx}
\caption{Exact finite-horizon expected corruption.}
\label{tab:appendix-2}
\end{table}
E1 is consistent with the scalar fixed point and geometric transient. Its validity is limited to the exact binary inheritance model. In particular, a real reviewer may both detect a bad input and introduce a new error, which is not represented by a catch-only level.

\section{E2: role topology and finite-population sampling}
For each K in \{2,3,5\}, E2 constructs three stochastic input matrices. In the chain topology, each role reads the preceding role (with the first reading itself). In the random topology, each row is sampled from independent positive weights and normalized. In the hub topology, non-hub roles read the hub and the hub reads role zero. Heterogeneous g and catch rates are spread linearly across roles.

The analytic prediction solves the linear system exactly. The finite system has independent agent outcomes conditional on the current empirical role means. We simulate 450 steps, discard the first 250, and average role fractions over the remaining 200 time points. Balanced populations use N\_k=4,000; imbalanced populations use N\_k=1,000(k+1).

\begin{table}[t]
\centering
\small
\setlength{\tabcolsep}{2pt}
\begin{tabularx}{\columnwidth}{>{\hsize=0.4000\hsize\raggedright\arraybackslash}X>{\hsize=1.1200\hsize\raggedright\arraybackslash}X>{\hsize=1.2000\hsize\raggedright\arraybackslash}X>{\hsize=1.2800\hsize\raggedright\arraybackslash}X}
\toprule
\textbf{K} & \textbf{Topology} & \textbf{Balanced} & \textbf{Imbalanced} \\
\midrule
2 & chain & 0.002 & 0.004 \\
2 & random & 0.001 & 0.002 \\
2 & hub & 0.000 & 0.001 \\
3 & chain & 0.001 & 0.005 \\
3 & random & 0.000 & 0.001 \\
3 & hub & 0.000 & 0.001 \\
5 & chain & 0.001 & 0.005 \\
5 & random & 0.001 & 0.001 \\
5 & hub & 0.000 & 0.001 \\
\bottomrule
\end{tabularx}
\caption{E2 error by topology (mean absolute role error).}
\label{tab:appendix-3}
\end{table}
\subsection{Results and interpretation}
Across the 60 role-level comparisons, the mean absolute discrepancy is 0.001537 and the maximum is 0.010379. The largest discrepancy occurs in the smallest imbalanced role population and is expected to be more variable than a balanced 4,000-agent role.

The experiment covers input topology and population imbalance, but not every possible allocation of verification across roles. In particular, it does not optimize the catch-rate vector subject to a fixed budget. Such an optimization would require an explicit role-weighted objective and is not implied by the fixed-point calculation.

Because the stochastic update uses a binomial count with probability computed from current role means, it is equivalent to independent source sampling with replacement for the empirical fraction. We do not simulate individual edges, which would add computational cost without changing this transition law.

\subsection{Failure modes tested}
The random topology is regenerated from the fixed experiment generator, so the exact matrix is recorded implicitly by the seed rather than exported in the current CSV. A future artifact release should serialize each matrix alongside the results to permit bitwise cross-language reproduction.

\section{E3: concentration rate}
The E3 population sizes are 250, 500, 1,000, 2,000, 4,000, 8,000, 16,000, 32,000, and 64,000 agents per role. At each size and each mixing rate, 100 independent trajectories are simulated for 35 time steps from an all-correct initial condition. We average absolute error over replicas for each role, then report the maximum of those role-wise means.

\begin{figure}[t]
\centering
\includegraphics[width=\columnwidth]{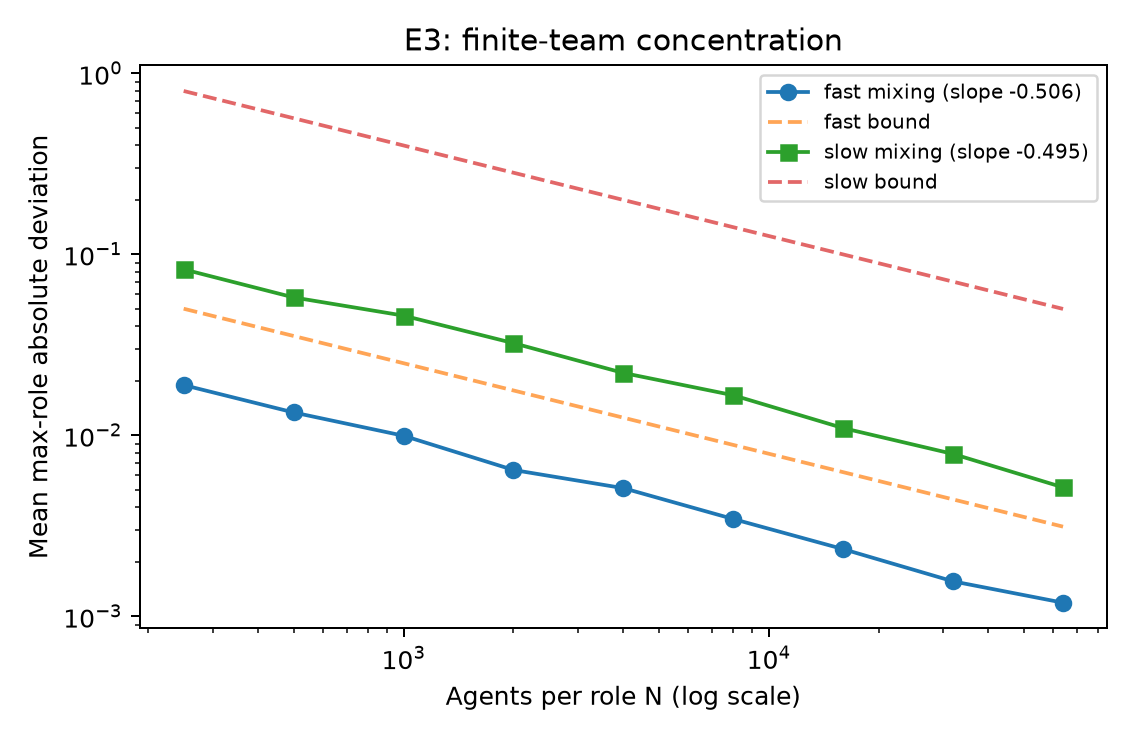}
\caption{Finite-population deviation and the theorem's conservative bound.}
\label{fig:paper-3}
\end{figure}
\begin{table}[t]
\centering
\small
\setlength{\tabcolsep}{2pt}
\begin{tabularx}{\columnwidth}{>{\hsize=0.9000\hsize\raggedright\arraybackslash}X>{\hsize=0.7200\hsize\raggedright\arraybackslash}X>{\hsize=0.7800\hsize\raggedright\arraybackslash}X>{\hsize=1.2000\hsize\raggedright\arraybackslash}X>{\hsize=1.2000\hsize\raggedright\arraybackslash}X>{\hsize=1.2000\hsize\raggedright\arraybackslash}X}
\toprule
\textbf{Mixing} & \textbf{u} & \textbf{Slope} & \textbf{Error at N=250} & \textbf{Error at N=64k} & \textbf{Mean bound violations} \\
\midrule
fast & 0.368 & -0.506 & 0.019 & 0.001 & 0 \\
slow & 0.960 & -0.495 & 0.083 & 0.005 & 0 \\
\bottomrule
\end{tabularx}
\caption{E3 fitted slopes and endpoint errors.}
\label{tab:appendix-4}
\end{table}
\subsection{Results}
The fitted slopes are -0.506 (fast) and -0.495 (slow), close to the predicted -1/2 rate. The slow-mixing setting has a larger finite-population bound because u is closer to one. All 18 mean-error summaries lie below their stated bound. This comparison is at the mean-summary level; a high-probability confidence bound is not claimed.

\subsection{Initial-condition ablation}
The main theorem compares the finite and mean-field systems from the same initial condition. The separate initial-condition CSV records a deterministic gap after 35 steps when the finite population starts fully corrupted and the mean field starts fully correct. The initial discrepancy is not covered by the same-initial-condition bound; it decays under contraction according to the trajectory difference.

The ablation is not folded into the reported slope because it changes the estimand. It highlights why initialization and transient horizon must be stated when using mean-field approximations for short projects or tasks.

\section{E4: filtering and policy structure}
The policy grid has 1,001 equally spaced beliefs from 0 to 1. At each belief, the script evaluates all four one-stage costs with g\_N=g\_D=0.08, delta=0.20, mu=0.35, lambda=10, kappa=(0,0.7,0.35,1.5), and the catch rates in the main text. The check asks whether level 2 occupies an initial interval and whether the action is constant after that interval.

The five signal settings p in \{0.30,0.50,0.70,0.90,0.99\} share q=0.02. Because the action does not affect the signal process and the optimal action is myopic, signal quality does not change the computed stage-cost minimizer as a function of belief. The interval ends at approximately b=0.345 and the tail action is level 3 in each grid.

\begin{table}[t]
\centering
\small
\setlength{\tabcolsep}{2pt}
\begin{tabularx}{\columnwidth}{>{\hsize=0.6000\hsize\raggedright\arraybackslash}X>{\hsize=0.6000\hsize\raggedright\arraybackslash}X>{\hsize=0.9000\hsize\raggedright\arraybackslash}X>{\hsize=1.3200\hsize\raggedright\arraybackslash}X>{\hsize=1.0800\hsize\raggedright\arraybackslash}X>{\hsize=1.5000\hsize\raggedright\arraybackslash}X}
\toprule
\textbf{p} & \textbf{q} & \textbf{Signal gap} & \textbf{Mean delay} & \textbf{Median} & \textbf{Policy form} \\
\midrule
0.30 & 0.02 & 0.28 & 5.065 & 3.00 & pass \\
0.50 & 0.02 & 0.48 & 2.535 & 2.00 & pass \\
0.70 & 0.02 & 0.68 & 1.470 & 1.00 & pass \\
0.90 & 0.02 & 0.88 & 1.250 & 1.00 & pass \\
0.99 & 0.02 & 0.97 & 1.030 & 1.00 & pass \\
\bottomrule
\end{tabularx}
\caption{E4 structural check and post-change delay.}
\label{tab:appendix-5}
\end{table}
\begin{figure}[t]
\centering
\includegraphics[width=\columnwidth]{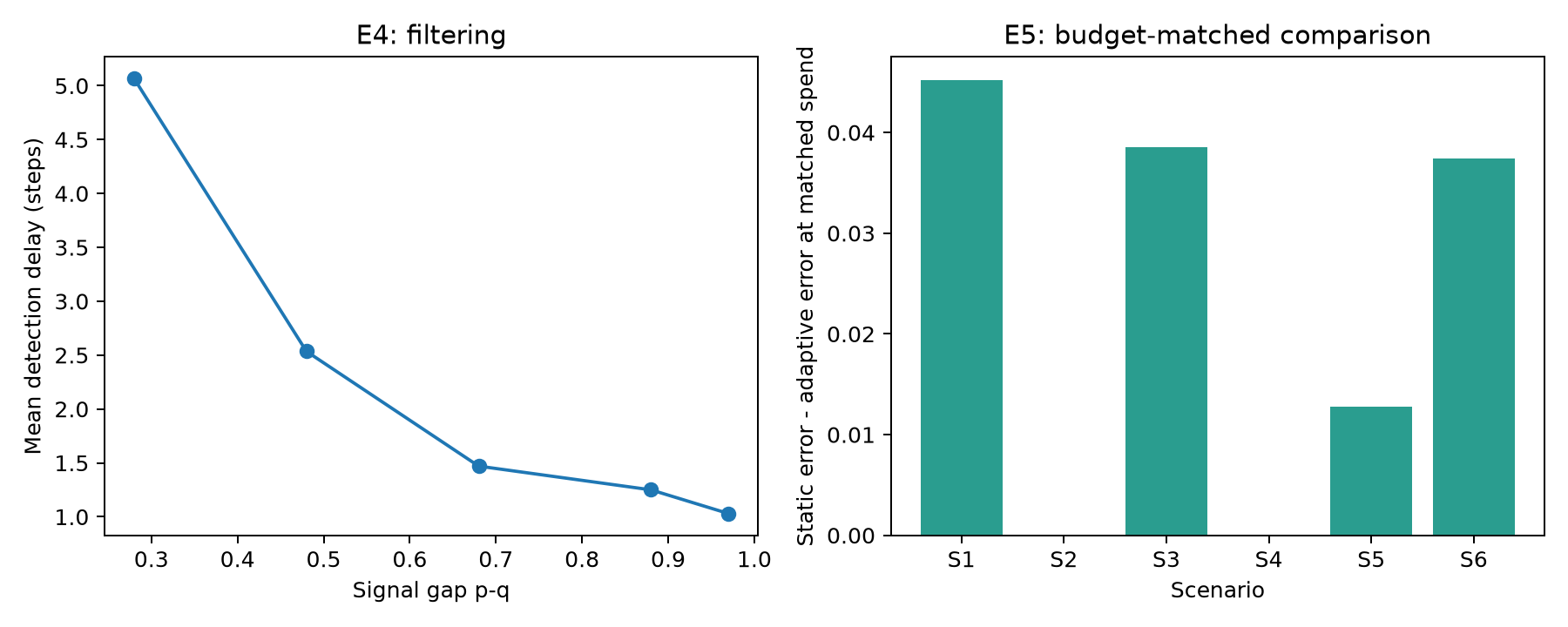}
\caption{Left: detection delay versus signal gap. Right: E5 equal-budget error differences.}
\label{fig:paper-4}
\end{figure}
\subsection{Detection-delay protocol}
For each p setting, 200 runs begin with the stationary prior for alpha=beta=0.03 and then receive only degraded-regime signals. Delay counts observations until the filtered belief first reaches 0.9, with a 1,000-step cap. This post-change protocol is intentionally conditional on a regime change; it does not estimate false-alarm rates or average run length under N.

Mean delay declines from 5.065 at p=0.30 to 1.030 at p=0.99. The reduction is monotone over the five tested settings, but the small number of signal-quality values and the fixed switching prior do not establish a general monotonicity theorem for detection delay.

\section{E4 ablations: switching and calibration}
The switching-rate ablation varies alpha=beta over 0.005, 0.03, and 0.15. For each, the true post-change signal parameters are p=0.90 and q=0.02. The filter is run once with the correct p=0.90 and once with a misspecified p-hat=0.70; q remains correctly specified. Each cell has 200 runs.

\begin{table}[t]
\centering
\small
\setlength{\tabcolsep}{2pt}
\begin{tabularx}{\columnwidth}{>{\hsize=1.2500\hsize\raggedright\arraybackslash}X>{\hsize=0.8500\hsize\raggedright\arraybackslash}X>{\hsize=1.0000\hsize\raggedright\arraybackslash}X>{\hsize=0.9000\hsize\raggedright\arraybackslash}X>{\hsize=1.0000\hsize\raggedright\arraybackslash}X}
\toprule
\textbf{alpha=beta} & \textbf{Filter p-hat} & \textbf{Mean delay} & \textbf{Median delay} & \textbf{90th pct.} \\
\midrule
0.005 & 0.70 & 1.170 & 1.00 & 2.00 \\
0.005 & 0.90 & 1.235 & 1.00 & 3.00 \\
0.030 & 0.70 & 1.130 & 1.00 & 2.00 \\
0.030 & 0.90 & 1.225 & 1.00 & 1.00 \\
0.150 & 0.70 & 1.110 & 1.00 & 2.00 \\
0.150 & 0.90 & 1.075 & 1.00 & 1.00 \\
\bottomrule
\end{tabularx}
\caption{E4 switching and model-calibration ablation.}
\label{tab:appendix-6}
\end{table}
\subsection{Reading the stress results}
The prior is initialized at alpha/(alpha+beta), then a D-regime signal stream is supplied. Under a misspecified filter, posterior updates use the wrong likelihood ratio. The measured delay therefore reflects both signal evidence and model calibration. Faster nominal switching changes the prediction step and can reduce the posterior's persistence.

These are conditional detection-delay summaries, not a properly calibrated statistical detection analysis. We do not report Type-I error, average run length, or a sequential test's false-alarm guarantee. Those quantities would be necessary for deployment decisions.

\subsection{Structural policy versus signal quality}
The policy structure is determined by the belief-to-cost map, not by p and q directly. A weaker signal makes the belief less informative over time, but the pointwise optimal action at a given b is unchanged. A future closed-loop controller may have signal quality enter the action through the continuation value; this paper's myopia result rules that out only under its exogenous-state assumptions.

Filter misspecification can still produce poor realized decisions even though the pointwise policy is optimal for the supplied belief. Robust filtering and online estimation of p, q, alpha, and beta are not addressed.

\section{E5: equal-budget policy comparison}
E5 simulates 300,000 time points for each scenario. The true regime follows a two-state chain with alpha=beta=0.03; signals are drawn conditionally from p and q. The controller filters beliefs and selects the action minimizing posterior expected one-step cost. Realized expected corruption is evaluated using the true regime's generation and catch parameters.

The baseline is the best static randomized mixture of the four levels with exactly the same mean verification spend as the adaptive policy. Because there is one spend constraint, the optimum over static mixtures can be represented by a mixture of at most two levels. The script enumerates all feasible pairs and linearly interpolates error at the adaptive spend.

\begin{table*}[t]
\centering
\scriptsize
\setlength{\tabcolsep}{2pt}
\begin{tabularx}{\textwidth}{>{\hsize=0.4500\hsize\raggedright\arraybackslash}X>{\hsize=1.1700\hsize\raggedright\arraybackslash}X>{\hsize=0.6300\hsize\raggedright\arraybackslash}X>{\hsize=0.5400\hsize\raggedright\arraybackslash}X>{\hsize=1.1700\hsize\raggedright\arraybackslash}X>{\hsize=0.8100\hsize\raggedright\arraybackslash}X>{\hsize=1.4400\hsize\raggedright\arraybackslash}X>{\hsize=1.4400\hsize\raggedright\arraybackslash}X>{\hsize=1.3500\hsize\raggedright\arraybackslash}X}
\toprule
\textbf{S} & \textbf{(p,q)} & \textbf{d} & \textbf{mu} & \textbf{Differ?} & \textbf{Cost} & \textbf{Adaptive} & \textbf{Static} & \textbf{Gain} \\
\midrule
S1 & 0.90/0.02 & 0.20 & 0.35 & 1 & 0.938 & 0.1373 & 0.1825 & 0.0452 \\
S2 & 0.02/0.02 & 0.20 & 0.35 & 1 & 1.500 & 0.1122 & 0.1122 & 0.0000 \\
S3 & 0.50/0.02 & 0.20 & 0.35 & 1 & 0.979 & 0.1393 & 0.1778 & 0.0385 \\
S4 & 0.90/0.02 & 0.80 & 0.35 & 0 & 0.350 & 0.1846 & 0.1846 & 0.0000 \\
S5 & 0.90/0.02 & 0.20 & 0.10 & 1 & 0.175 & 0.1386 & 0.1514 & 0.0128 \\
S6 & 0.90/0.02 & 0.20 & 0.35 & 1 & 0.937 & 0.1757 & 0.2131 & 0.0374 \\
\bottomrule
\end{tabularx}
\caption{E5 errors at matched verification spend (d is runner degradation).}
\label{tab:appendix-7}
\end{table*}
\subsection{Main comparison}
For S1 (p=0.90, q=0.02, delta=0.20, mu=0.35; g\_N and g\_D are both 0.08), adaptive mean error is 0.1373 at spend 0.938. The best static mixture gives error 0.1825 (gain 0.0452). At p=q=0.02, both errors are 0.1122 and the measured gain is zero.

When delta=0.80, the same runner level is optimal in both regimes; the adaptive policy and matched static mixture coincide with zero measured gain. With mu=0.10, a smaller but positive gain of 0.0128 is observed; with regime-dependent generation failure (0.04,0.20), gain is 0.0374.

The comparison supports the value-of-information intuition within the chosen parameterization. It is not a robust policy evaluation: there is no held-out task set, uncertainty interval, or empirical model calibration.

\section{E5 ablations: input corruption and correlated failures}
The E5 sensitivity scenarios vary signal informativeness, runner fragility delta, inherited-error probability mu, and the generation-failure gap between regimes. These factors have distinct roles: signal quality controls belief calibration; delta controls how much level 2 loses under degradation; mu controls the value of catching an inherited error; and g\_D-g\_N changes error independently of tool quality.

The uninformative-signal scenario sets p=q exactly. Its posterior stays at the stationary prior, so the adaptive controller selects one constant level. The exact zero matched-budget gain in this case is a useful implementation check. The delta=0.80 scenario instead has informative signals but a common regime-optimal level; it also yields zero gain in this sampled configuration.

\subsection{Correlated failures}
A population stress introduces a shared generation shock with probability 0, 0.02, or 0.05 and adjusts the idiosyncratic probability to preserve marginal g=0.10. Two thousand clusters of eight agents are followed for 300 steps at catch rate 0.55; the final-half mean and within-cluster generation-failure correlation are recorded.

\begin{table}[t]
\centering
\small
\setlength{\tabcolsep}{2pt}
\begin{tabularx}{\columnwidth}{>{\hsize=0.7200\hsize\raggedright\arraybackslash}X>{\hsize=0.9600\hsize\raggedright\arraybackslash}X>{\hsize=1.0800\hsize\raggedright\arraybackslash}X>{\hsize=1.2400\hsize\raggedright\arraybackslash}X}
\toprule
\textbf{Common shock} & \textbf{Empirical corruption} & \textbf{Independent fixed point} & \textbf{Within-cluster corr.} \\
\midrule
0.00 & 0.1684 & 0.1681 & -0.003 \\
0.02 & 0.1675 & 0.1681 & 0.173 \\
0.05 & 0.1672 & 0.1681 & 0.468 \\
\bottomrule
\end{tabularx}
\caption{E5 shared-shock population stress.}
\label{tab:appendix-8}
\end{table}
The shock is shared by all eight agents in a cluster at each time step, so generation failures are genuinely correlated within clusters. The marginal failure probability is held fixed, but the independent-agent concentration result does not apply. The experiment is descriptive and is not a replacement theorem.

\subsection{Budget baseline details}
Static mixes are randomized independently over units and have no access to the regime signal. Verification spend is the level's specified kappa; error is the scenario's expected corrupted-unit share. We compare errors only at the adaptive policy's observed average spend. The displayed gain is static error minus adaptive error, so positive values favor adaptation.

The baseline does not include a static policy that can condition on role identity, task type, or historical observations. Such policies would be stronger comparators and could narrow the reported differences.

\section{E6: randomized structural checks}
E6 draws 400 two-level parameter sets and directly compares the analytic endpoint condition with brute-force monotonicity of h on a 2,001-point belief grid. It also checks that the minimum-cost two-action policy is nondecreasing when the analytic condition holds. The randomized check agrees in 400/400 cases; all 389/389 cases satisfying the condition have a nondecreasing grid policy.

A separate 100-instance four-level draw with g\_N=g\_D checks the initial-level-2/constant-tail structure. All 100/100 instances satisfy the grid property. Randomized checks support but do not prove the proposition; the analytic argument is the proof.

\subsection{General-generation counterexamples}
We then draw 500 four-level systems with g\_D>g\_N and evaluate the optimal action at 2,001 beliefs. A failure is counted if the action sequence cannot be represented as level 2 on one initial interval followed by a constant tail. 9 instances fail under the fixed seed, a rate of 1.8\%. This is a counterexample search, not an estimate of the probability of failure under any real parameter distribution.

\begin{table}[t]
\centering
\small
\setlength{\tabcolsep}{2pt}
\begin{tabularx}{\columnwidth}{>{\hsize=1.0400\hsize\raggedright\arraybackslash}X>{\hsize=0.6000\hsize\raggedright\arraybackslash}X>{\hsize=1.0000\hsize\raggedright\arraybackslash}X>{\hsize=1.3600\hsize\raggedright\arraybackslash}X}
\toprule
\textbf{Check} & \textbf{Trials} & \textbf{Pass/failure} & \textbf{Interpretation} \\
\midrule
Two-level derivative condition & 400 & 400/400 agreement & grid corroboration \\
Threshold under condition & 389 & 389/389 threshold & pointwise policy \\
Four-level, equal g & 100 & 100/100 structure & restricted proposition \\
Four-level, g\_D>g\_N & 500 & 9 counterexamples & general claim fails \\
\bottomrule
\end{tabularx}
\caption{E6 randomized checks.}
\label{tab:appendix-9}
\end{table}
The counterexamples show why the four-level result is restricted to equal generation rates; no generic structure is asserted.

\section{Additional limitations and reporting notes}
The experiments use one master seed. Replication counts within E1, E3, E4, and E5 estimate Monte Carlo variation conditional on the selected parameters, but they do not replace independent seed-level replication. The E6 random parameter sample is generated from a hand-selected distribution and should not be read as a prior over deployment systems.

The E1 ablations use independent chain paths; the E2 and E3 population transitions use binomial role counts rather than agent-level graph objects. These implementations preserve the intended transition probabilities, but cross-checking against a second independent implementation would improve confidence.

The paper's results are narrow in another sense: the input corruption probability mu in the allocation cost is treated as exogenous. In a self-consistent team, mu depends on the policy and the network fixed point. Jointly solving that dependence could change the budget frontier and policy thresholds.

\subsection{Conclusion}
This work provides an analytically tractable bridge between error propagation in role-structured teams and verification allocation under hidden tool degradation. Within the model, the serial fixed point, affine mean-field equilibrium, finite-population rate, myopic policy, and restricted threshold structure are exact under their assumptions. Synthetic experiments check those predictions; E7 is only a small model pilot, not external validation.

\textbf{L. Exploratory hosted-LLM pilot.} Nine hosted model requests (four paired baseline/self-check tasks and one batch of eight arithmetic claims) yielded 4/4 correct in each task condition and verifier TP=5, FP=0, FN=0, TN=3. Auto routing selected two model variants. The two generated count\_runs implementations each passed five restricted unit tests. Total measured usage was 9 premium requests, 799,499,500 NanoAIU, 116,734 input tokens, 1,017 output tokens, and 167.5 seconds wall time. This tiny pilot has no tool-degradation treatment and does not evaluate a multi-agent team or the proposed allocation policy.

The nine per-call records sum exactly to the reported request, NanoAIU, input-token, and output-token totals. Their API durations sum to 29.567 seconds; 167.5 seconds is end-to-end process wall time.

\textbf{Data and code.} Synthetic results and privacy-reviewed pilot prompts, responses, and per-call usage records are in the public repository and source archive. No participant data were used.

\FloatBarrier
\end{document}